\documentclass[reprint,showpacs,preprintnumbers,superscriptaddress,multibbl,amsmath,amssymb,prl]{revtex4-2}

\usepackage[english]{babel}
\usepackage[utf8]{inputenc}
\usepackage{graphicx,wasysym}
\usepackage{dcolumn}
\usepackage{bm}
\usepackage[colorinlistoftodos, color=green!40, prependcaption]{todonotes}
\usepackage[pdftex, pdftitle={Article}, pdfauthor={Author}]{hyperref} 
\usepackage{amsthm}
\usepackage{mathtools}
\usepackage{physics}
\usepackage{xcolor}
\usepackage{graphicx}
\usepackage[left=23mm,right=13mm,top=35mm,columnsep=15pt]{geometry} 
\usepackage{adjustbox}
\usepackage{placeins}
\usepackage[T1]{fontenc}
\usepackage{lipsum}
\usepackage{csquotes}

\newcommand{\bx}{{\boldsymbol x}}

\newcommand{\btau}{{\boldsymbol \tau}}

\newcommand{\reals}{\mathbb R}

\newcommand{\E}{\mathbb E}

\begin{document}

\title{Extremal events dictate population growth rate inference}

\author{Trevor GrandPre}
\thanks{These authors contributed equally to this work.}
\affiliation{Department of Physics, Princeton University, Princeton, NJ 08544, USA}
\affiliation{Princeton Center for Theoretical Science, Princeton University, Princeton, NJ 08544, USA}
\affiliation{Center for the Physics of Biological Function, Princeton University, Princeton, NJ 08544, USA}
\author{Ethan Levien}
\thanks{These authors contributed equally to this work.}
\affiliation{Department of Mathematics, Dartmouth College, Hanover, NH 03755, USA}
\author{Ariel Amir}
\affiliation{Department of Physics of Complex Systems, Weizmann Institute of Science, Rehovot 76100, Israel}

\begin{abstract}
Recent methods have been developed to map single-cell lineage statistics to population growth. Because population growth selects for exponentially rare phenotypes, these methods inherently depend on sampling large deviations from finite data, which introduces systematic errors. A comprehensive understanding of these errors in the context of finite data remains elusive. To address this gap, we study the error in growth rate estimates across different models. We show that under the usual bias-variance decomposition, the bias can be decomposed into a finite-time bias and nonlinear averaging bias.   We demonstrate that finite-time bias, which dominates at short times, can be mitigated by fitting its monotonic behavior. In contrast, at longer times, nonlinear averaging bias becomes the predominant source of error, leading to a phase transition. This transition can be understood through the Random Energy Model, a mean-field model of disordered systems, where a few lineages dominate the estimator. Applying these methods to experimental data demonstrates that correcting for biases in lineage-based approaches yields consistent results for the long-term growth rate across multiple methods and enables the reverse-engineering of dynamic models. This new framework provides a quantitative understanding of growth rate estimators, clarifies the conditions under which they can be effectively applied to finite data, and introduces model-free approaches for studying the connections between physiology and cell growth.
\end{abstract}


\setcounter{secnumdepth}{6}
\maketitle
% \tableofcontents

\section{Introduction}
A central objective in biology is to understand the relationship between genotype, phenotype, and fitness \cite{orr2009fitness,de2014empirical,cadart2019physics,yamauchi2022unified,nozoe2017inferring,jafarpour2022evolutionary,lin2020evolution,lambert2015quantifying,fink2023microbial,genthon2020fluctuation}. 
In the context of microbes, a key component of fitness is the long-term growth rate \cite{lin2020evolution}, defined as
\begin{equation}
\Lambda = \lim_{T \to \infty}\frac{1}{T} \ln \left[\frac{N(T)}{N(0)}\right],
\end{equation}
where \(N(0)\) and \(N(t)\) are the population sizes at times \(t=0\) and \(t=T\). 

Experimentally, $\Lambda$ can be measured by bulk fitness assays. When performed on libraries of different strains, such experiments can yield insight into the genotype-to-fitness map \cite{kinsler2020fitness}. 
However, they do not reveal how individual single-cell traits contribute to fitness. The theoretical foundations of this question can be traced back to seminal work in demography, dating to Euler \cite{bacaer2011short}. Most notably, the Euler-Lotka equation relates the population growth rate to the distribution of individual lifetimes within a population. For a population undergoing binary fission, this relationship is expressed as ~\cite{powell1956growth,lotka1907relation,bacaer2011short,smith1977estimates}
\begin{equation}\label{eq:el}
   \E[e^{-\Lambda \tau}] = \frac{1}{2}~.
\end{equation}
In its modern form \cite{levien2020interplay,lebowitz1974theory,lin2017effects}, which accommodates correlated generation times, the average, $\E[\cdot]$, is taken over all generation times  \(\tau\) throughout the entire population tree. A key implication of Eq.~\ref{eq:el} is that 
\begin{equation}
\label{jensen}
    \Lambda \ge \frac{\ln(2)}{\E[\tau]}
\end{equation}
which follows from Jensen’s inequality. In a population where every cell divides after a fixed time $\tau_0$, the growth rate is exactly $\ln(2)/\tau_0$. Equation (3) implies that introducing variability in generation times, while keeping the same mean generation time, cannot decrease the population growth rate below this bound. However, variability in single-cell growth rates, while fixing the mean growth rate,  may lower the population growth rate compared to a population without such variability~\cite{lin2017effects, lin2020single}. In this case, the inequality in Eq. (3) still holds, but the impact of growth rate fluctuations is captured through changes in $\E[\tau]$, leading to a slower population growth rate.

\begin{figure*}
\begin{center}
\includegraphics[width=0.95\textwidth]{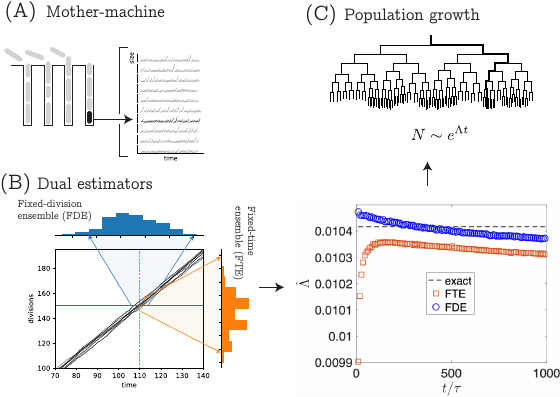}
\caption{(A) Schematic of the mother-machine microfluidic device used to collect single-cell lineage data, alongside an example of lineage data. (B) Dual ensembles can be used to obtain large deviation statistics for predicting the growth rate. The distribution for the blue histogram is obtained by taking a horizontal cross-section of the data at 150 divisions, while the orange is obtained by taking a vertical cross-section.
By using these histograms to estimate the scaled cumulant generating function, we obtain the fixed-divisions ensemble (FDE) as shown in Eq.~\eqref{eq:fde} where the divisions are fixed and the time durations vary and fixed-time ensemble (FTE) as shown in Eq.~\eqref{eq:Lambdapsi} where the total time duration is fixed and the number of divisions vary as defined in Section \ref{sec:estimators}. (C) These lead to estimators of the exponential growth rate of a growing population whose lineages, when sampled by travelling forward in time from the root of the tree, have the same division statistics as the lineages in (A). The plot below the population tree shows the estimators applied to a model of cell-size control (see Appendix~\ref{sec:csc}). The dotted line is the exact answer and the red and blue squares are from the FTE and the FDE, respectively. }
\label{Fi:Fig1}
\end{center}
\end{figure*}

Mother machine experiments offer a powerful investigative tool to explore the effects of mutations on single-cell lineage dynamics (see Fig.~\ref{Fi:Fig1}A). In these experiments, cells are confined in microfluidic channels, with all but one lineage being expelled from the experiment in each channel~\cite{thiermann2024tools,tanouchi2017long}. Barcoding techniques have been developed to enable mother machine assays across multiple strains simultaneously \cite{camsund2020time}. A key question is how to connect the data from these experiments to data from bulk fitness assays. Although the Euler-Lotka equation provides a theoretical framework for connecting single-cell generation time statistics to bulk fitness, the average in Eq.~\ref{eq:el} is taken over the entire population tree, whereas mother machines capture data from only single lineages. These two approaches coincide only when generation times are uncorrelated.

Recently, studies have explored the challenges in predicting the population growth rate from single-cell lineage statistics \cite{levien2020large,lin2020single,pigolotti2021generalized,nozoe2017inferring,genthon2020fluctuation}. One estimator of the population growth rate from lineages is described in Ref.~\cite{levien2020large}:
\begin{equation}
\label{oldeq}
\hat{\Lambda}_{1}=\frac{1}{T}\ln\left[\frac{1}{M}\sum_{i=1}^M 2^{N_t^{(i)}}\right]~
\end{equation}
where $N_t^{(i)}$ is the number of divisions by a fixed time $T$ and $M$ is the number of lineages (see Sec.~\ref{sec:estimators} for more details). In practice, this method is sensitive to extremal statistics of the sampled generation times or division counts, leading to non-monotonic convergence in the duration of time intervals.  At short times, this estimator has a monotonic convergence with time. At longer times (and fixed $M$), the sample averages become dominated by extremal statistics. As a result, the naive estimate on the right-hand side of Eq.\eqref{jensen}~ is obtained rather than the true growth rate \cite{levien2020large}.

Another estimator of the population growth rate from lineages is described in Ref~\cite{pigolotti2021generalized}:
\begin{equation}
\label{ELO}
\frac{1}{n}\ln\left[\frac{1}{M}\sum_{i=1}^{M}e^{-T_{n}^{(i)}\hat{\Lambda}_{2}}\right]=-\ln(2)~.
\end{equation}
Here, $T_{n}^{(i)}$ represents the time of each lineage at a fixed number of divisions  $n$, and we can solve for  $\hat{\Lambda}_{2}$  in the exponent. Unlike the standard Euler-Lotka equation in Eq.~\eqref{eq:el}, this formulation is a generalized Euler-Lotka equation tailored to observables along lineages, making it particularly suitable for mother-machine data.

If we have an infinite amount of data, $\hat{\Lambda}_{1}=\hat{\Lambda}_{2}$. However, our understanding of how these two methods perform in practice with finite data is incomplete. In this paper, we address this gap by demonstrating that, within the conventional bias-variance trade-off framework~\cite{demidenko2019advanced, breiman1996bias}, the bias can be decomposed into two distinct components: a finite-time bias that persists even with an infinite number of lineages, and a nonlinear averaging bias arising when the lineage ensemble is not self-averaging (Section \ref{sec:estimators}). In Section \ref{sec:ft}, we show that finite-size scaling—a well-established concept in statistical physics~\cite{goldenfeld2018lectures}—can be employed to completely eliminate the finite-time bias.

In Section \ref{sec:ft}, we establish that finite-lineage effects can be understood through their connection to the Random Energy Model (REM), a mean-field model of disordered systems. This connection clarifies how non-monotonic convergence arises from a phase transition into a ``frozen state''. In the REM, this transition corresponds to the system entering a low-temperature phase dominated by a few energy states. Similarly, in the context of growth-rate estimators, this transition occurs when a few extremal lineages begin to dominate the ensemble, leading to analogous behavior. Our analysis draws parallels with previous studies on Jarzynski estimators of free energy differences \cite{suarez2012phase,palassini2011improving,rohwer2015convergence}, and also connects to research on estimating large deviations, particularly in the context of predicting the bandwidth of tele-traffic streams \cite{lewis1998practical,duffy2015estimating,patch2020ranking}. We explore these connections further in Appendix \ref{sec:otherwork}. Our findings provide a framework for determining when growth rate estimation is feasible from single-cell data.

\section{Dual estimators and error decomposition}\label{sec:estimators}
We consider the general setting in which single-cell generation times evolve stochastically according to some process $\{\tau_k\}$. This need not be a Markov process, but to be concrete we imagine there is some underlying phenotype (e.g. gene expression) $x_k \in {\mathbb R}^d$ which evolves according to a Markov process with transition operator $h(x_{k+1}|x_k)$, and that generation times are deterministic functions of the phenotype $\tau(x)$. This modeling framework can capture all existing models of single-cell dynamics \cite{levien2021non,amir2014cell,ho2018modeling}.  We let $T_n = \sum_{k=1}^n\tau_k$ denote the  time at which the $n$th cell in a lineage divides. 

The long-term growth rate is related to the lineage-to-lineage fluctuations in the counting process,
\begin{equation}
N_t = \max\left\{n: \sum_{i=1}^n\tau_i < t\right\}~,
\end{equation}
and  $\Lambda$ is given by~\cite{levien2020large}
\begin{equation}\label{eq:Lambdapsi}
\Lambda = \psi(\ln(2))
\end{equation}
 where $\psi(z)$ is the scaled-cumulant generating function,
\begin{equation}
\psi(z) = \lim_{t \to \infty}\frac{1}{t}\ln \E[e^{N_tz}]~.
\end{equation}
Here, $z$ is the conjugate variable to $N_t$, and \(\E[\cdot]\) represents the expected value with respect to the lineage distribution within the \textit{fixed-time ensemble (FTE)}, as illustrated in Fig. \ref{Fi:Fig1}B and C. The intuition behind Eq.~\eqref{eq:Lambdapsi} is that lineages with $n$ divisions on average contribute $2^{N_t}$ cells to the final population, hence the total population size is on average $\E[2^{N_t}] \sim e^{\Lambda t}$.

Given lineage samples $\{N_t^{(i)}\}$ which come from repeated experiments or splitting a long lineage into blocks, $\psi(z)$ and hence $\Lambda$ can in principle be estimated by replacing the expectation with an empirical average: 
\begin{equation}\label{eq:fte_scgf}
\hat{\psi}_t(z)  =  \frac{1}{t}\ln \left\langle e^{N_t^{(i)}z} \right\rangle_M 
\end{equation}
where $\langle \cdot \rangle_M$ is the empirical average over $M$ samples. 
An estimator of $\Lambda$ is then $\hat{\Lambda}_1(t,M) = \hat{\psi}_t(\ln(2))$ which is equivalent to Eq.~\eqref{oldeq}, and is always a biased estimator.  

These formulas naturally connect to the large deviations of \(N_t\) through the Gärtner-Ellis Theorem, which states that
\begin{equation}\label{eq:Ntldp}
-\lim_{t \to \infty}\frac{1}{t} \ln P(N_t/t = x) = I(x),
\end{equation}
where \(I(x)\) is the \emph{large deviation rate function}, defined as the Legendre-Fenchel transform \(\psi^*(x)\) of \(\psi(x)\). Informally, Eq. \ref{eq:Ntldp} tells us that the fluctuations in \(N_t/t\) decay exponentially with time: \(P(N_t/t = x) \approx A_t e^{-t I(x)}\), where $A_t$ is the normalization constant.

In Ref.~\cite{levien2020large}, it was shown that due to the large deviation structure, the estimator \(\hat{\Lambda}_1\) exhibits a somewhat surprising non-monotonic convergence. This phenomenon is related to the so-called linearization effect~\cite{rohwer2015convergence}, which can be understood as follows. The integral \(\E[ e^{xz}] \approx \int p_t(x)e^{xz t}\, dx\) is dominated by a value \(x^*=N^*_t/t\) for which \(\psi(z)=x^*z - I(x^*)\) is extremal. Therefore, to obtain an accurate estimate of \(\E[ e^{N_tz}]\) from finite samples, we must have a high probability of sampling \(N_t = N_t^*\). However, when \(t\) is large, this is an exponentially rare event, requiring an exponentially large number of samples. 

As shown in Ref.~\cite{pigolotti2021generalized}, the growth rate can alternatively be expressed in terms of the scaled cumulant generating function (SCGF) of \(T_n\), \(\phi(z) = \lim_{n \to \infty} n^{-1} \ln \E[e^{T_n z}]\), as
\begin{equation}\label{eq:fde_exact}
-\phi(-\Lambda) = \ln(2).
\end{equation}
This result arises from the fact that \(\phi(z) = \psi^{-1}(z)\) for the SCGFs of a counting process and its dual first-passage time process~\cite{gingrich2017fundamental}, which we refer to as the \textit{fixed-divisions ensemble (FDE)} \footnote{Note that in Ref.~\cite{pigolotti2021generalized}, this estimator was called the Generalized Euler-Lotka (GEL) Equation.}. A comparison of these ensembles is shown in Fig.~\ref{Fi:Fig1}B and C. Additionally, the large deviation rate function for \(T_n/n\), denoted by \(J\), can be related to \(I\) through the expression \(J(y) = xI(1/x)\)~\cite{gingrich2017fundamental}.

Following Ref.~\cite{pigolotti2021generalized} and Eq.~\eqref{ELO}, an estimator $\hat{\Lambda}_2$ can be obtained as the positive solution to the nonlinear equation
\begin{equation}\label{eq:fde}
- \hat{\phi}_n(-\hat{\Lambda}_2(n,M)) = \ln(2),
\end{equation}
where \(\hat{\phi}_n(z)\) is the empirical SCGF for the dual process:
\begin{equation}\label{eq:fde_scgf}
 \hat{\phi}_n(z) = \frac{1}{n} \ln \left\langle e^{T_n^{(i)} z} \right\rangle_M.
\end{equation}
Since \(\hat{\phi}_n(z)\) is convex,  \(\hat{\Lambda}_2(n,M)\) is uniquely defined by Eq.~\eqref{eq:fde}.

Our primary goal is to analyze the convergence behavior of the FTE and FDE ensembles, denoted by \(\hat{\Lambda}_1\) and \(\hat{\Lambda}_2\), respectively, as a function of the parameters $M$ and $t$ (for FTE) and  $M$ and $n$ (for FDE). Although the dual estimator \(\hat{\Lambda}_2\) also appears to have systematic errors from the linearization effect, it is unclear whether one estimator consistently outperforms the other or how their convergence patterns are influenced by specific model details.

\subsection{Error decomposition}
For both the FTE and FDE, we define the error to be
\begin{equation}
\label{errort}
\varepsilon(\hat{\Lambda}) = \sqrt{\E[(\hat{\Lambda}-\Lambda)^2]}.
\end{equation}
Assuming perfect measurements of division times and counts, applying the standard bias-variance decomposition yields \cite{demidenko2019advanced, breiman1996bias}:

\begin{equation}
\begin{split}
\label{error1}
\varepsilon^2(\hat{\Lambda}) & = \underbrace{\E[(\hat{\Lambda}- \E[\hat{\Lambda}])^2]}_{\coloneqq{\rm var}(\hat{\Lambda})}+ \underbrace{ (\E[\hat{\Lambda}]- \Lambda)^2}_{\coloneqq{\rm Bias}(\hat{\Lambda})^2}
 \end{split}
\end{equation}
where $\E[\cdot]$ is an ensemble average of $\hat{\Lambda}$ from many realizations of the growth process. This formula follows from the definition of ${\rm var}$ and $\varepsilon(\hat{\Lambda})$.

To fully grasp the non-monotonic convergence, it is crucial to closely examine the bias term, which can be further decomposed as:
\begin{align}
\label{error2}
{\rm Bias}(t,M) = \underbrace{(\E[\hat{\Lambda}]- \Lambda(t))}_{{\rm Bias}_{{\rm nl}}(t,M)}+ \underbrace{(\Lambda(t)- \Lambda)}_{{\rm Bias}_{\rm ft}(t)}
\end{align}
where $\Lambda(t) = t^{-1}\ln \E[2^{N_t}]$ and $\Lambda=\lim_{t\rightarrow \infty} \Lambda(t)$. Note that by taking the expectation, we assume the large $M$ limit has been reached, with fixed lineage durations or division counts for the FTE and FDE, respectively. However, in order to compute $\varepsilon(\hat{\Lambda})$ from Eq. \eqref{errort} and its different contributions from Eq. \eqref{error1} using simulations, we must replace all averages by their sample averages as detailed in Eqs. \eqref{eq:fte_scgf} and \eqref{eq:fde_scgf} over many realizations of the growth process. 

The nonlinear averaging bias, ${\rm Bias}_{\rm nl}(t,M)$, is intimately connected to the concept of \emph{quenched free energy} in disordered systems, a relationship that emerges when $\langle \cdot \rangle_M$ is interpreted as a partition function. This connection will be clarified in Section \ref{sec:finite_lineage}, where we introduce a scaling of lineage durations with $M$, analogous to the approach used in \cite{suarez2012phase,palassini2011improving} to estimate free energy differences. It is important to note that the dependence on $M$ arises solely from ${\rm var}(\hat{\Lambda})$ and ${\rm Bias}_{\rm nl}$, and both go to zero in the limit that $M$ goes to infinity. Hence, the only error at large $M$ is the finite-time error:
\begin{equation}
\lim_{M \to \infty}\varepsilon^2(\hat{\Lambda}) = {\rm Bias}_{\rm ft}(t)
\end{equation}

In the next section, we will present numerical and theoretical results and explore the differences in the finite-time bias between the FTE and FDE.

\subsection{Numerical results}\label{sec:num}
We now present numerical results that highlight the key features of the different terms in the error expression. We conduct simulations using a simple model where generation times follow a first-order autoregressive process (AR1)~\cite{lin2020single, lin2017effects,cerulus2016noise}:
\begin{equation}
\label{modeleq1}
\tau_{n+1} = \tau_{0}(1-c) + \tau_{n}c + \xi_n,
\end{equation}
where $\tau_{0}$ is the average generation time, $\tau_n$ is the parent generation time, $c$ is the Pearson correlation coefficient between parent and offspring, and $\xi_n$ represents noise with zero mean and a variance given by $\sigma_{\xi}^2 = (1-c^2)\sigma_{\tau}^2$. 
This model incorporates correlations between generation times, which are essential for maintaining homeostasis, and captures key aspects of the convergence for the FTE and FDE estimators. 

The long-time population growth rate for the model in Eq.~\eqref{modeleq1} was first calculated in Ref.~\cite{lin2020single} to be
\begin{equation}\label{eq:FDE_c}
  \Lambda =\frac{2 \ln(2)/ \tau_0}{1 + \sqrt{1-2 \ln(2)(\sigma_T/\tau_0)^2}}
\end{equation} 
where 
\begin{equation}
\sigma_{T}^2=\sigma_{\tau}^2\frac{(1+c)}{(1-c)}~.
\end{equation}

An alternative derivation of the population growth rate of this model using a \textit{Second Cumulant Expansion (SCE)} of FDE is discussed in Appendix~\ref{sec:cltmethod}.

In Fig.~\ref{Fi:Fig2}A, we show the total error for FTE and FDE from  Eq.~\eqref{errort} from simulations over \(t/\tau_0  = 10^3\) generations. The parameters used were \(\tau_0 = 1\) and \(\sigma_\tau = 0.2\), with 10, 20, and 40 lineages. Note that the error decreases over time for both methods, but the FDE has an order of magnitude smaller error for the first 100 generations. The methods have comparable error after the first 100 generations.

Next, we look at the different contributions to error. In Figs. \ref{Fi:Fig2}B and C, we show the four contributions to the total error based on Eq.~\eqref{error1} for FTE and FDE, respectively. At short times, the FTE estimator is dominated by the finite-time bias, while the FDE estimator is dominated by the variance of the estimator, though it retains a small, nonzero finite-time bias. For both methods, the variance of the estimators tends to zero as the number of generations tends to infinity, making the total error primarily driven by the nonlinear bias, \({\rm Bias}_{\rm nl}(t,M)\).

\begin{figure*}[t!]
\begin{center}
\includegraphics[width=16cm]{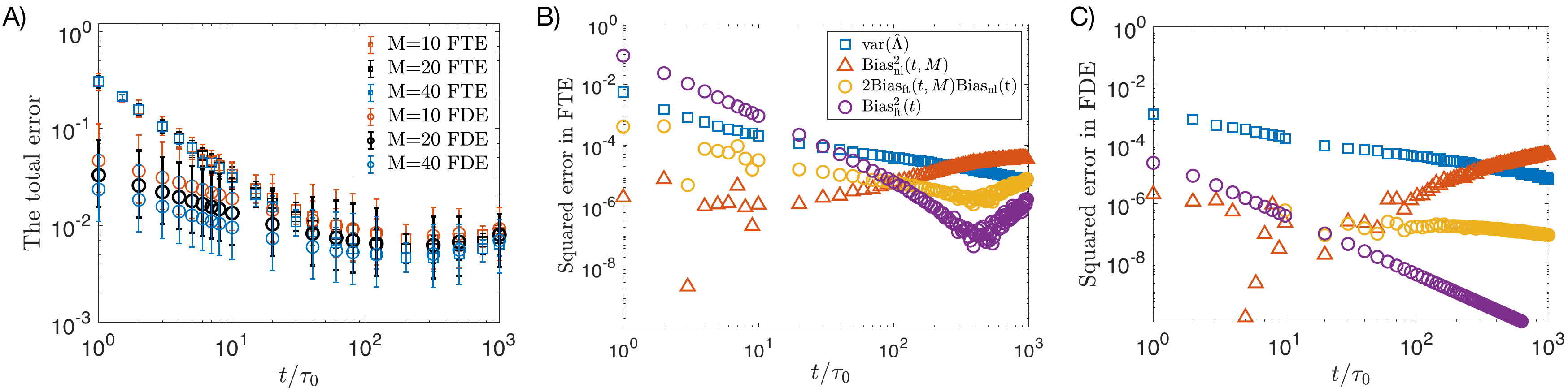}
\caption{Systematic error in growth rate estimates. Simulations were done using the model in Eq.~\eqref{modeleq1} with parameters given in Sec. \ref{sec:num}. The analytical solution of the long-time population growth rate of this model is shown in Eq.~\ref{eq:FDE_c}. (A) The average total error for both estimators (Eq.~\eqref{errort}) for $M=10$, $M=20$, and $M=40$ lineages over $10^3$ generations, calculated from \(10^3\) realizations, with error bars representing the standard deviation. (B) The absolute squared error (Eqs.~\eqref{error1} and \eqref{error2}) and all contributions for FTE. (C) The absolute squared error (Eqs.~\eqref{error1} and \eqref{error2}) and all contributions for FDE.  }
\label{Fi:Fig2}
\end{center} 
\end{figure*}

In the next section, we will discuss the monotonic convergence of ${\rm Bias}_{\rm ft}(t)$ for both methods and why the FDE has a much smaller error than FTE. Then, in sec. \ref{sec:finite_lineage} we will discuss the nonlinear averaging bias, ${\rm Bias}_{\rm nl}(t,M)$.

\section{Finite duration bias (${\rm Bias}_{\rm ft}(t)$)}\label{sec:ft}
In this section, we examine the behavior of the finite time and finite division number error in both the FTE and FDE. Specifically, we demonstrate that both ensembles exhibit inverse scaling with lineage duration given that the lineage number is large enough, while providing a justification for the significantly smaller prefactor observed in the FDE.

\subsection{Inverse time scaling}
In previous work~\cite{levien2020large}, we demonstrated that for small noise in generation times, $\hat{\Lambda}_1$ converges inversely with lineage length, as described by
\begin{equation}
\label{finiteLIFT}
    {\rm Bias}_{\rm ft}(t,M) = \frac{A}{t}+o\left(\frac{1}{t}\right)~.
\end{equation}
Generally, deriving an exact expression for the finite-time bias is challenging, even when the large deviations are well understood. However, as we discuss below, in certain cases, this term can be approximately removed from the estimator using a finite-time scaling approach.

Next, we demonstrate that a similar result can be obtained for the FDE estimator as follows: We can express the average on the left-hand side of Eq.~\eqref{eq:fde} in terms of the large deviation function,
\begin{equation}
\label{gel1}
    \left\langle e^{-T_{n}^{(i)}\hat{\Lambda}_2} \right\rangle = \frac{K_n}{n} \int e^{-n t \hat{\Lambda}_2 - n I(t)}\, dt~,
\end{equation}
where \(t = T/n\) and \(K_n\) is a normalization constant. We can compute the convergence of the FDE estimator exactly for all times for the model in Eq. \eqref{modeleq1} (see Appendix \ref{ap1} for derivation). The leading order correction to the FDE will be
\begin{equation}
\label{finileg}
    \hat{\Lambda}_2 = \Lambda + \frac{B}{n}+o\left(\frac{1}{n}\right)~,
\end{equation}
where $B$ is the finite-time coefficient. In practice, the coefficients for both FDE and FTE are determined directly from the data by fitting the coefficients \(A\) and \(B\) at short times.

We demonstrate this in Fig. \ref{Fi:Fig3} for the model in Eq. \eqref{modeleq1} and show the fitting procedure and the finite-time coefficients as a function of the correlation strength, $c$. The finite-time coefficients are obtained by plotting the total error vs. $\tau_0 /t$ and extracting the slope at small times. In this format, fitting the linearly decreasing points from right to left gives the finite-time coefficients. As the number of generations gets too large, the nonlinear averaging bias dominates and the error becomes non-convex and begins to increase again. At the point that the error begins to increase, we stop the fitting. If there were an infinite amount of lineages, the estimators would linearly decrease until zero.

In the inset of Fig.~\ref{Fi:Fig3}, we show the value of the finite time coefficients over a range of $c$ values. We find that the coefficient for the FDE is consistently about 10 times smaller than that of the FTE. In the next subsection, we will show why the FDE has such a smaller coefficient.

\subsection{Transport equation derivation of FDE}
\label{relativecoefficients}
We now present an alternative derivation of the FDE estimator using the von Foerster equation approach. This method has the advantage of clarifying the numerical results discussed in the previous section. In essence, the significantly smaller bias of the FDE estimator arises from the absence of finite-time bias, provided that generation times are uncorrelated across generations ($c=0$ within the model shown in Eq. \eqref{modeleq1}).

To understand this more precisely, recall that we assume the Markov transition operator for the generation time dynamics, \(h(x|y)\), is such that \(x_k\) converges to a unique stationary distribution \(\rho(x)\) for any initial \(x_0\)~\cite{levien2020interplay}. Given that \(N(a,x,t)\) represents the number of cells with age \(a\) and phenotype \(x\) at time \(t\), the von Foerster equation is expressed as  
\begin{align}
\frac{\partial}{\partial t}N(a,x,t) &= -\frac{\partial}{\partial a}N(a,x,t),\quad a<\tau(x) \\
N(0,x,t) &= 2 \int_{0}^{\infty} h(x|x') N(\tau(x'),x',t)\, dx'. 
\end{align}
(This form is slightly different from most standard definitions, where the phenotype \(x\) is not considered.) The total number of cells at time \(t\) is given by \(N(t) = \int_{0}^{\infty} \int_0^a N(a,x,t)\, da\, dx\), and given our assumptions on \(x\), for large \(t\), \(\phi(a,x,t) = N(a,x,t)/N(t)\) converges to a steady-state \(\varrho(x,\tau)\) that satisfies  
\begin{equation}
 -\frac{\partial}{\partial a}\varrho(a,x) = -\Lambda \varrho(a,x) \implies \varrho(a,x) = \varrho(0,x)e^{-\tau(x)\Lambda}~.
\end{equation}
This leads to the equation 
\begin{align}\label{eq:phi0}
\begin{split}
\varrho(0,x) &= 2 \int h(x|x')\varrho(x')\, dx' \\
&= 2 \int h(x|x') \varrho(0,x')e^{-\tau(x')\Lambda}\, dx'~.  
\end{split}
\end{align}
Integrating over \(x\): 
\begin{equation}\label{eq:el2}
1 = 2 \int \varrho_b(x)e^{-\tau(x)\Lambda}\, dx 
\end{equation}
where \(\varrho_b(x)\) is the "birth distribution" of \(x\). Equation \eqref{eq:el2} is a generalization of the well-known Euler-Lotka equation.

Note that in Equation \eqref{eq:el}, \(\varrho_b(x)\) must be obtained from the population distribution. To derive a relation that depends solely on lineage statistics, we iteratively replace \(\varrho(0,x')\) using Eq. \eqref{eq:phi0}, yielding
\begin{align}\label{eq:phi0m}
\begin{split}
\varrho_b(x_n) &= 2^n \int \cdots \int \varrho_b(x_0)\prod_{i=1}^n h(x_i|x_{i-1}) e^{-\tau(x_{i-1})\Lambda} \, dx_i \, dx_0 \\
&= 2^n \int \cdots \int \varrho_b(x_0) e^{-T_n\Lambda} \prod_{i=1}^n h(x_i|x_{i-1}) \, dx_i \, dx_0~,\\
\end{split}
\end{align}
where \(T_n = \sum_{i=1}^{n} \tau(x_i)\) is the total time along a lineage of \(n\) divisions. 

Integrating over \(x_n\), taking logarithms, and dividing by \(n\) gives
\begin{equation}\label{eq:fde2}
-\ln 2 =  \frac{1}{n} \ln \E_{T,\varrho_b}\left[ e^{-T_n\Lambda} \right], 
\end{equation}
where \(\E_{T,\varrho_b}\) denotes the expectation of \(T_n\), with the first cell drawn from \(\varrho_b\).

Equation \eqref{eq:fde2} holds for all \(n\), but if we replace \(\varrho_b\) with another distribution, \(\tilde{\varrho}\), we must consider the limit of a large number of divisions to derive an equivalent form of Eq.~\eqref{eq:fde_exact}:
\begin{equation}\label{eq:fde1}
- \ln 2 = \lim_{n \to \infty} \frac{1}{n} \ln \E_{T,\tilde{\varrho}}\left[ e^{-T\Lambda} \right].
\end{equation}
This derivation is distinct from that in Ref.~\cite{pigolotti2021generalized} and has important implications for the estimator in Eq.~\eqref{eq:fde}. The finite-time bias arises only from the discrepancy between the initial distribution \(\tilde{\varrho}\), assumed to be the lineage distribution, and the population distribution \(\varrho_b\). Consequently, this bias vanishes when there are no correlations (i.e., when \(h(x|y) = h(x)\)) for FDE. This contrasts with the FTE estimator~\cite{levien2020large}, where there is always a finite-time correction regardless of the initial distribution or correlations between generation times.
\begin{figure}[t!]
\begin{center}
\includegraphics[width=8.0cm]{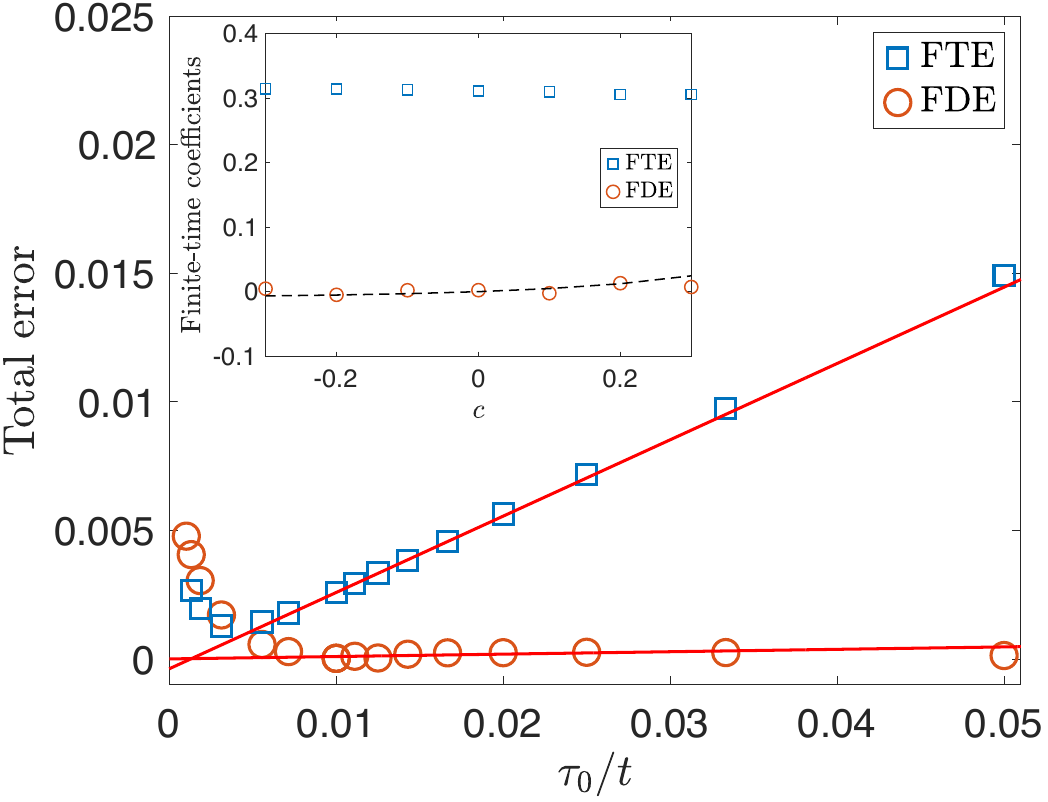}
\caption{The finite-time coefficients. The linear trend from right to left (red lines) is due to the finite-time error (see Eqs.~\eqref{finiteLIFT} and \eqref{finileg}). The blue squares and red circles represent data from the FTE and FDE, respectively, obtained from 1,000 simulations using the same parameters as Fig.~\ref{Fi:Fig2}. Error bars are omitted because they are significantly smaller than the symbols. For FDE, the x-axis represents the number of divisions,  $n = \tau_0 / t$. The inset shows the finite-time coefficients for a range of correlation coefficients. The black dotted line is the prediction of the FDE from Eq.~\eqref{exactcoeff}. At larger times, the total error is no longer linearly decreasing in time so the fit is stopped when the error becomes nonconvex. }
\label{Fi:Fig3}
\end{center} 
\end{figure}

As shown in Fig.~\ref{Fi:Fig3}, there is a finite-time bias for both FTE and FDE, which decreases monotonically with time initially. At longer times, however, the total error begins to increase nonlinearly. This behavior arises because the nonlinear averaging bias starts to dominate the total error. The phenomenon of non-monotonic convergence for FTE was first identified in Ref.~\cite{levien2020large}, where it was observed that, for large times and a fixed number of lineages, the estimator approaches the naive estimate on the right-hand side of Eq.~\eqref{jensen}. However, previous work did not fully characterize the transition from monotonic to non-monotonic convergence in the total error for FTE, and this transition has not been observed before for FDE.

Next, we demonstrate that the crossover from monotonic to non-monotonic convergence in the total error represents a second-order phase transition. Furthermore, we quantitatively show how this systematic error can be avoided.
\section{Nonlinear averaging bias $({\rm Bias}_{\rm nl})$ and connection to the Random Energy Model}\label{sec:finite_lineage}

Here we discuss the bias resulting from the finite number $M$ of lineages. Inspired by the approach in Refs.~\cite{suarez2012phase,palassini2011improving}, we obtain an approximate expression for the bias in the FDE using a known formula for the free energy density of the REM.

As discussed in Sec.~\ref{sec:estimators} and demonstrated by our numerical results in Sec.~\ref{sec:num}, the empirical averages used to estimate the SCGF are influenced by a linearization effect, a phenomenon described in Ref.~\cite{rohwer2015convergence}. This effect is analogous to the error seen in Jarzynski's Equality estimators of free energy differences \cite{suarez2012phase}, which can be understood through a connection to the Random Energy Model (REM). The REM is a mean-field model for disordered systems, which assumes that the energies of each state are sampled from an independent and identically distributed (iid) Gaussian variable~\cite{derrida1980random}. In this model, the partition function \( Z_N \) is simply an iid sum:
\begin{equation}
Z_N = \sum_{i=1}^{2^N} e^{-\beta \sqrt{N/2} X_i},
\end{equation}
where \( X_i \) are iid standard normal variables which would represent energy states in REM, but will represent individual lineages in our context. Additionally, within the original REM, $N$ and $\beta$ would be the system size and the inverse temperature. As we show below, in our case the $N$ is related to the logarithm of the number of lineages and $\beta$ is related to the ratio of the logarithm of the number of lineages to the number of divisions. Consequently, many properties of the REM can be derived using classical extreme value theory, without resorting to the more complex mathematical tools often required in the study of disordered systems~\footnote{In many statistical mechanics papers, \( X_i \) are taken to have a variance of \(1/2\). Therefore, we have introduced a factor of \(1/2\) in the scaling of \(N\) to align our definition of the inverse temperature with the standard literature.}. 

We focus on the asymptotic behavior of the free energy density, defined as
\begin{equation}\label{eq:fed}
    f_N(\beta) = -\frac{1}{\beta N} \ln Z_N,
\end{equation}
with the limit of the quenched average given by $\bar{f}(\beta) = \lim_{N \to \infty} \E[f_N(\beta)]$. In the thermodynamic limit, we find the free energy to be~\cite{derrida1980random}:
\begin{equation}\label{eq:fed_limit}
\bar{f}(\beta) = \begin{cases} 
- \frac{\beta}{4} -\frac{\ln 2}{\beta}, & \text{if } \beta \leq \beta_c, \\ 
-\sqrt{\ln 2}, & \text{if } \beta > \beta_c.
\end{cases}
\end{equation}
where $\beta_c = 2\sqrt{\ln 2}$ is the critical inverse temperature. The free energy in the high temperature regime $\beta < \beta_c$ is entropically dominated and can be obtained by replacing $Z_N$ with $\E[Z_N]$ in Eq. \eqref{eq:fed}, avoiding the need to evaluate the quenched average. This is due to a concentration of the Gibbs measure, which is well established for the REM.  In contrast, for the low temperature regime $\beta \ge \beta_c$ the free energy density is determined by the extremal energy levels. It can be shown that the transition corresponds to the breakdown of the Law of Large Numbers for $Z_N$ \cite{ben2005limit}. Within this phase, the partition function is dominated by a few energy levels. As we show below, within our context the growth rate estimators are dominated by a few extremal lineages past the phase transition.

We begin by examining the connection of REM to the FDE, which offers two key advantages. First, (i) in this context, we can approximate $T_n^{(i)}$ using a Gaussian distribution, whereas for the FTE, we must contend with the counting variables $N_t^{(i)}$. Second, (ii) the finite-time bias in the FDE is minimal and can be effectively eliminated in our simulations by setting the mother-daughter correlations to zero, thereby allowing us to isolate ${\rm Bias}_{\rm nl}$.

The goal in this section is to understand when the system will not converge to the correct population growth rate due to the nonlinear averaging bias. We use the AR1 process defined in Eq.~\eqref{modeleq1} for the simulations in this section. We express the exponent of Eq. \eqref{eq:fde_scgf} as
\begin{equation}
zT_n^{(i)} = z \tau_0 n + z \sigma \sqrt{n} X_i,
\end{equation}
where $X_i$ follows a standard normal distribution~\footnote{It is important to note that this expression is valid only when the coefficient of variation, ${CV}_T$, is sufficiently small, ensuring that the likelihood of negative times remains negligible.}. Next, we set $M = 2^N$ and fix
\begin{equation}
\label{alphav}
\alpha_{\rm FDE} = \sqrt{\frac{2n}{N}}.
\end{equation}
Equation \eqref{eq:fde_scgf} can then be rewritten in terms of the (finite-size) free energy density of the REM by introducing the temperature parameter \(\beta_{\rm FDE}(z) = z \sigma \alpha_{\rm FDE}\):
\begin{align}\label{eq:fde_scgf_rem}
\lim_{n \to \infty}\E[\hat{\phi}_n(-z)] &= -\frac{\beta_{\rm FDE}(z)}{\alpha^2_{\rm FDE}}f_N(\beta_{\rm FDE}(z)) \nonumber \\
&\quad - \frac{\ln(2)}{\alpha^2_{\rm FDE}}- z\tau_0.
\end{align}

Equation \eqref{eq:fde_scgf_rem} is an exact equation but we can't solve for $\hat{\Lambda}$ analytically, since we don't have a closed formula for the \emph{finite-size} free energy density $f_N(\beta(z))$. However, we can study the large $N$ behavior by replacing $f_N(\beta(z))$ with the formula for $\bar{f}(\beta(z))$ given in Eq.~\eqref{eq:fed_limit}. The resulting equation can be solved to yield an estimate of the FDE estimator $E[\hat{\Lambda}] \approx \hat{\Lambda}_{\rm REM}(\alpha_{\rm FDE})$ which has the explicit formula
\begin{equation}\label{eq:Lhat_rem}
 \hat{\Lambda}_{\rm REM}(\alpha)=  \left\{ \begin{array}{lr}
 \Lambda & \alpha < \alpha_c\\
 \frac{2\ln(2)(1-\alpha^2/2)}{\tau_0 (2 \alpha\sigma/\tau_0 \sqrt{\ln(2)} - \alpha^2)} &\alpha \ge \alpha_c
 \end{array}\right.
\end{equation}
with the critical value of $\alpha$ given by
\begin{equation}
\label{critical_alpha}
  \alpha_c =   \frac{2\sqrt{\ln(2)}}{\Lambda \sigma }~.
\end{equation}
Here $\Lambda$ is given by Eq.~\eqref{eq:FDE_c}. Note that a useful approximation to $\alpha_c$ is 
\begin{equation}
\alpha_c \approx 2/(\sqrt{\ln(2)}{CV}_{T})~.
\end{equation}
It can be checked that $\hat{\Lambda}_{\rm REM}(\alpha)$ is indeed continuous at $\alpha_c$ and as $\alpha \to \infty$ tends to $\ln(2)/\tau_0$ since the transition to the frozen state is second order. 

In Fig. 4, we show the phase transition as a function of $\alpha_{\rm FDE}$ for the FDE from simulations with $c=0$, $\tau=1$, and $\sigma=0.1$. Small $\alpha_{\rm FDE}$ corresponds to the high temperature regime and large $\alpha$ would be analogous to a small temperature regime which is predicted by Eq.~\eqref{critical_alpha}. As the finite-time bias vanishes with $n \to \infty$, Eq.~\eqref{critical_alpha} provides insight into when ${\rm Bias}_{\rm nl}$ begins to increase, and thereby how small we need $\alpha$ to be in order to obtain reliable growth rate estimates. Indeed, our numerical experiments demonstrate that the REM-based theory effectively captures the transition in ${\rm Bias}_{\rm nl}$. In addition, the error bars decrease as $\alpha$ increases because the variance between realizations—described by the first term on the right-hand side of Eq.~\ref{error1}—monotonically decreases over time. In Appendix \ref{REM_A}, we show that, similar to the FDE, the FTE also quantitatively agrees with the REM framework. However, the decay after the phase transition differs slightly between the two ensembles.

\begin{figure}[h!]
\begin{center}
\includegraphics[width=8.5cm]{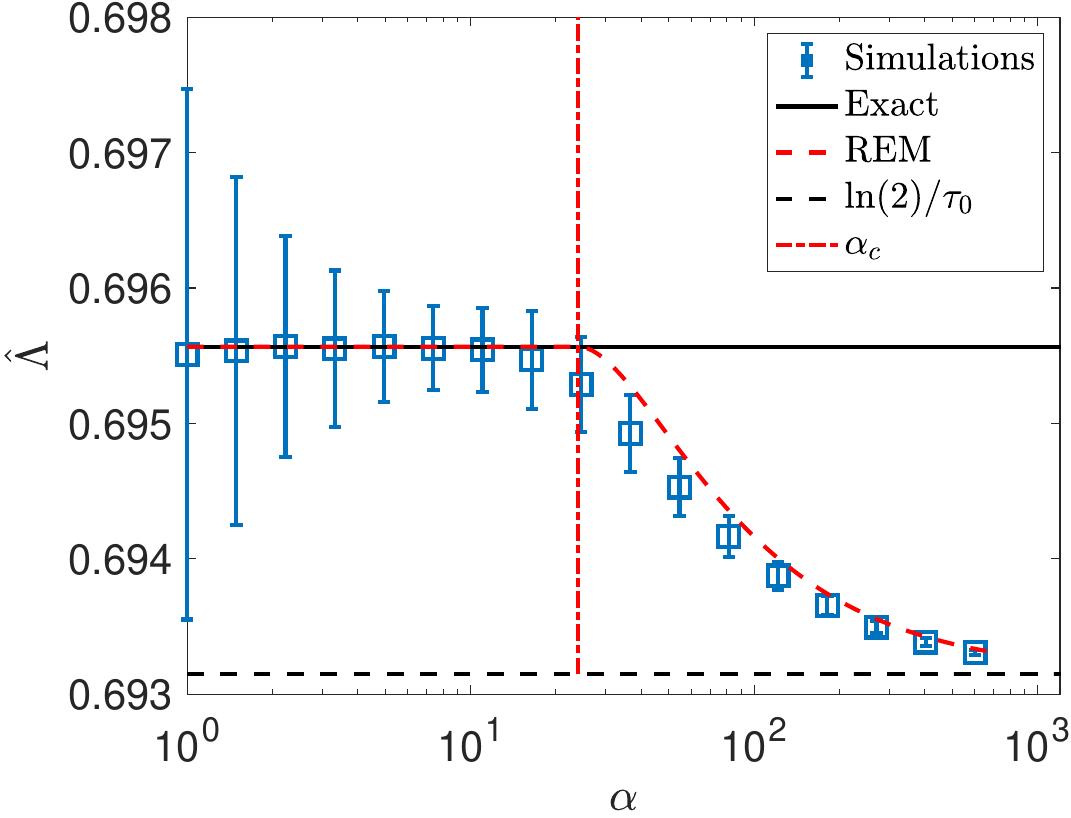}
\caption{Estimates of $\Lambda$ using the FDE compared to the theoretical prediction of the Random Energy Model (Eq. \eqref{eq:Lhat_rem}). The blue squares represent the mean of 1,900 realizations, with error bars indicating the standard deviation. These error bars monotonically decrease as $\alpha$ increases, due to the monotonically decreasing variance term in the first term on the right-hand side of Eq.~\ref{error1}. Simulations are performed with independent Gaussian generation times with $\sigma = 0.1$ and $c=0$. Different values of $\alpha$ were realized by fixing $M$ and modulating $n$. The black solid line is the exact population growth rate. The red dotted line is the prediction from Eq.~\ref{eq:Lhat_rem}, the dotted black line is $\ln(2)/\tau_0$, and the vertical red dashed-dotted line represents $\alpha_c$
(Eq.~\eqref{critical_alpha}).}
\label{Fi:Fig_rem}
\end{center} 
\end{figure}

However, it is important to note that our definition of ${\rm Bias}_{\rm nl}$ is based on $\E[\hat{\Lambda}]$, where the expectation is taken after solving Eq.~\eqref{eq:fde_scgf}. In contrast, to derive  Eq.~\eqref{eq:Lhat_rem}, we employed Eq.~\eqref{eq:fed_limit}, where the expectation is taken before solving  Eq.~\eqref{eq:fde_scgf_rem}. 
Namely, the ordering of limits is different which could lead to systematic differences. Most likely, the deviations from the exact value seen in Fig.~\ref{Fi:Fig_rem} are due to finite-size effects. For a given $\alpha$, both $n$ and $\ln(M)$ should go to infinity while leaving the ratio $n/\ln M$ fixed. Deviations from REM were also seen in sampling Jarzynski's Equality~\cite{suarez2012phase}.

Using the equation for the $\alpha_c$ in Eq.~\eqref{critical_alpha}, we find that for $c=0.0$ and $\sigma=0.1$, $\alpha_c\approx 20$. For a given value for $M$, we can find the value of time durations before the nonlinear averaging bias takes over from the inequality:

\begin{equation}
n<\frac{\ln(M)\alpha_c^2}{2\ln(2)}~.
\end{equation}

For instance, when M=20 the transition is predicted to occur at about 1000 divisions. For smaller values of $M$, this transition will occur for smaller numbers of divisions as shown in the equation for $\alpha$ in Eq.~\eqref{alphav}.

\section{Application of estimators on real data}

Now that we have an understanding of the rate of convergence for both methods, we can apply this to experimental data to understand the $ {\rm Bias}_{\rm ft}$ which would correspond to the analysis in Sec.~\ref{sec:ft}. We consider \textit{E. coli} grown at 25 \textdegree C with data from Ref.~\cite{tanouchi2017long}. Given that there are 70 lineages, the methods are not expected to be affected by the nonlinear averaging bias. This is supported by the fact that $\alpha_{\rm FDE} \approx 5$, which is well below the critical threshold $\alpha_c \approx 100$, and the same holds for FTE.

\begin{figure}[h!]
\begin{center}
\includegraphics[width=8.8cm]{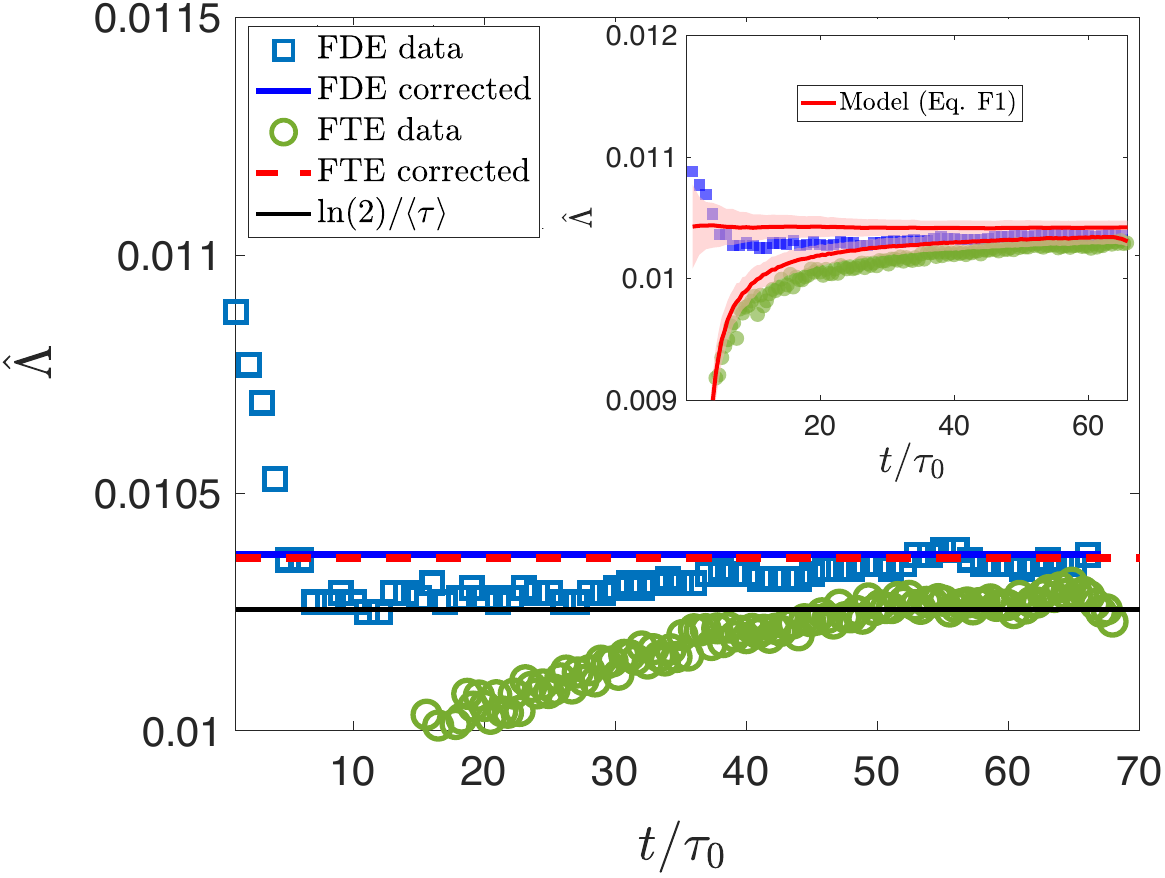}
\caption{Estimates of $\Lambda$ using FDE and FTE with finite size corrections. The symbols represent the uncorrected FDE (blue squares) and FTE (green circles). The red dotted line and blue solid line correspond to the finite-time corrected population growth rates for FTE and FDE, respectively, which are both approximately $0.0104 \ \text{min}^{-1}$. In the inset, we show that the cell-size control model described in Eqs.~\eqref{model1} and \eqref{model2}, which was fit to the data correlations, quantitatively reproduces the transient behavior observed in both methods. The symbols are the same data as FDE (blue squares) and FTE data (gree circles). The solid red lines represent the mean of the estimators over 100 realizations, while the shaded regions denote their standard deviation.}
\label{Fi:Fig4}
\end{center} 
\end{figure}

In Fig.~\ref{Fi:Fig4}, we show the performance of both FTE and the FDE on the data. Both methods asymptotically converge from below. The best fit lines for both methods are shown in Fig.~\ref{Fi:SI1}. When the two methods are fit and their finite-time error subtracted, the long-time growth rate estimates agree as shown in the red dotted (FTE) and blue (FDE) lines in Fig.~\ref{Fi:Fig4}. The long-time population growth rate estimate for the FDE is approximately $1.04 \times 10^{-2} \pm 1.97 \times 10^{-5} \ \text{min}^{-1}$ for FDE and $1.04 \times 10^{-2} \pm 3.11 \times 10^{-5} \ \text{min}^{-1}$ for FTE, which are both distinct from the naive estimate, $\ln(2)/\tau_0\approx 1.025 \times 10^{-2} \ \text{min}^{-1}$. Since the two estimates agree, we have some confidence that the true population growth rate is near this value. Additionally, the finite time coefficient for FTE ($A=-0.0060$) is about four times larger than FDE ($B=-0.0016$), in agreement with our theory in Sec.~\ref{sec:ft}.

Most of our analysis has focused on understanding the convergence of model-free estimators of the growth rate. However, these estimators of the growth rate can also be used to reverse-engineer a model for the dynamics, albeit with a reconstruction that might not be unique. In the inset of Fig.~\ref{Fi:Fig4} we show that a two-component model for the generation time and log size dynamics that are fit to the correlations of the data quantitatively reproduces the convergence of both methods (see Appendix~\ref{modelfit} for model details).

\section{Conclusions}
We have investigated the scaling behavior of errors in two estimators of population growth rate derived from lineage statistics. This problem is very similar to the estimation of free energy differences using Jarzynski's Equality \cite{suarez2012phase,palassini2011improving} and to the methods for controlling buffer size in ATM networks \cite{lewis1998practical} (see Appendix \ref{sec:otherwork}). These problems require sampling rare events, leading to a breakdown of traditional approaches to quantifying the sample distribution. Instead, it is essential to carefully account for extreme value statistics.

An effect of extremal statistics is the introduction of systematic error. Unlike the usual errors in statistical estimators, which stem from the lack of flexibility in the underlying model, this error arises because of poor sampling of the tails. In the context of growth rate estimations, there is an added complication which is the finite-time bias. We have demonstrated that there is, in fact, a trade-off between two types of bias: short-lineage lengths introduce a finite-time bias, while long-lineage lengths result in what we term a nonlinear averaging bias. This latter bias becomes significant when a few lineages dominate the sample averages $\langle \cdot \rangle_M$.

From the bias-variance decomposition of the total error, we find that at short times the finite length bias and the variance between realizations dominate the total error. At long times, there is a phase transition in which the linearization bias dominates. While the finite length bias can be mitigated through finite time-scaling, addressing the nonlinear averaging bias requires careful selection of lineage durations. By drawing a connection to the REM, we estimate the critical value of $\alpha^2=n\ln(2)/\ln M$ for FDE and $\alpha^2=t\ln(2)/\ln M$ for FTE at which the nonlinear averaging bias becomes dominant. This insight could be valuable in designing experiments that map single-cell data to population growth rates and has broader applications in contexts where one is interested in estimating large deviations of a counting process and its first passage time.

Regarding the connection to REM, it is intriguing to explore more rigorously whether the limiting behavior of $\hat{\Lambda}$ can be understood using similar techniques. It is well-known (see, e.g., \cite{bovier2006statistical,ben2005limit}) that the REM exhibits two phase transitions at $\beta_2 = \sqrt{\ln 2}$ and $\beta_1 = 2\sqrt{\ln 2}$. For $\beta > \beta_2$, the partition function $Z_N$ fails to obey the Law of Large Numbers, meaning that $Z_N/E[Z_N]$ no longer converges to one in probability. For $\beta > \beta_1$, the Central Limit Theorem no longer holds, implying that $(Z_N - E[Z_N])/\text{var}(Z_N)$ does not exhibit Gaussian fluctuations. A deeper understanding of these phase transition behaviors in $\hat{\Lambda}$ could enable more accurate inference in the future.

Lastly, understanding the biases and their trade-offs is important for applications involving Jarzynski’s Equality and ATM networks (see Appendix~\ref{sec:otherwork}). Moreover, it is interesting to explore the role of bias trade-offs more broadly in thermodynamic inference problems such as estimating entropy and entropy production rate (EPR)~\cite{strong1998entropy,roldan2010estimating,grandpre2024direct}
as well as numerical methods such as cloning to compute large deviation functions~~\cite{hidalgo2017finite,ray2018importance,grandpre2021entropy,angeli2019rare}. Additionally, there are some EPR methods that involve waiting time distributions~\cite{van2023time,pietzonka2024thermodynamic}, which have similarities to generation times. A precise understanding of biases inherent to finite data will allow accurate inference of fundamental properties.

\section*{Acknowledgments}
\begin{acknowledgments}
We would like to thank William Kath, David Lacoste, Hugo Touchette, and Ned Wingreen for useful discussions. 
 This work was supported in part by the National Science Foundation, through the Center for the Physics of Biological Function (PHY-1734030). T.G. was supported by the Schmidt Science Fellowship. A.A. was supported by the European Union (ERC, BIGR, 101125981), Israeli Science Foundation (146873) and the Clore Center for Biological Physics.
\end{acknowledgments}

\appendix

\section{Cell-size control model}\label{sec:csc}
In Fig.~\ref{Fi:Fig1}, we show simulations of a biophysically grounded model of microbial growth, commonly used to explore mechanisms of cellular size homeostasis 
\cite{ho2018modeling,amir2014cell}

This model presumes that individual cells grow exponentially until they reach a division size \( v_{\rm d} \), which is a function of their size at birth, \( v_{\rm b} \). For exponentially growing cells, the generation time \( \tau \) follows \( v_{\rm d} = v_{\rm b} e^{\lambda \tau} \), or equivalently,  
\begin{equation}
\tau = \frac{1}{\lambda} \ln \left( \frac{v_{\rm d}}{v_{\rm b}} \right),
\end{equation}
where \( \lambda \) represents the growth rate at the single-cell level. Assuming symmetric cell division, the size at birth, \( v_{\rm b} \), is half the size of the mother cell at division. To incorporate phenotypic variability, the model includes random fluctuations in both growth rates and division volumes. Specifically, each cell's growth rate \( \lambda \) and division volume \( v_{\rm div} \) are modeled by \cite{lin2020single}  

\begin{align}
	\ln \lambda &= \ln \lambda_0 + \eta_{\lambda}, \\
	v_{\rm div} &= 2(1 - \gamma) v_{\rm birth} + 2 \alpha v_0 + \eta_{v},
\end{align}
where \( \eta_{\lambda} \) and \( \eta_{v} \) are independent normal random variables with variances \( \sigma_{\lambda}^2 \) and \( \sigma_{v}^2 \), respectively. The growth rates follow a log-normal distribution with small noise, and the division timing is based on cell volume. Here, \( \alpha \) dictates the cell-size regulation approach: for \( \gamma= 1 \) (a “sizer” strategy), cells divide at a target size, whereas for \( \gamma = 1/2 \) (an “adder” strategy), cells increase by a consistent size \( v_0 \) from birth to division. 

\section{Derivation of the population growth rate of the simulation model SCE}\label{sec:cltmethod}
If the large deviation rate function is quadratic (e.g. $T_n$ is Gaussian), then we can do a cumulant expansion of Eq. \eqref{eq:fde_scgf} and truncate at second order. Starting with the series expansion
\begin{equation}
\ln \E[e^{T_n z}] 
= \sum_{k=1}^{\infty}\kappa_k \frac{z^k}{k!},
\end{equation}
and dropping all but the first two terms, and substituting into Eq.\,\eqref{eq:fde_exact} yields
\begin{equation}
\tau_0\Lambda - \frac{\sigma^2_T}{2}\Lambda^2 = \ln(2)
\end{equation}
where
\begin{align}
\tau_0 &= \lim_{n \to \infty}\frac{\E[T_n]}{n}\\
\sigma_T^2 &=  \lim_{n \to \infty}\frac{{\rm var}(T_n)}{n}.
\end{align}
Solving the quadratic equation for $\Lambda$ yields Eq.~\eqref{eq:FDE_c}. Since this formula is exact when the large deviation rate function is quadratic, and thus the statistics of $T_n$ are Gaussian, we call it the \textit{Second Cumulant Expansion (SCE)}. However, for real data, the accuracy of this approximation is not known a priori.

To illustrate the limitations of the SCE, we present a counter example where the method does not work. We examine a model in which generations are independent and $\tau = t_0$ with probability $1/2$, and $\tau = 1$ otherwise. Because this is an independent generation time model, the tree and lineage distributions are identical, resulting in the same equation from both the Euler-Lotka and FDE:
\begin{equation}
 e^{t_0 \Lambda} = e^{(t_0 - 1)\Lambda} + 1~
 \label{recursive}
\end{equation}
which can be numerically solved for $\Lambda$.

Now, we can determine the growth rate for the SCE by substituting
\begin{equation}
\tau_0 = \frac{t_0 + 1}{2}
\end{equation}
and
\begin{equation}
\sigma^2 = \frac{(t_0 - 1)^2}{4}
\end{equation}
into Eq.~\eqref{eq:FDE_c}, yielding
\begin{equation}
\label{SCEbad}
\Lambda = \frac{\dfrac{4 \ln(2)}{t_0 + 1}}{1 + \sqrt{1 - 2 \ln(2) \left(\dfrac{t_0 - 1}{t_0 + 1}\right)^2}}.
\end{equation}
Note that the SCE simplifies to the naïve estimate $\ln(2)/\tau_0$ when $t_0 = 1$, reflecting the zero-noise limit.

\begin{figure}[h!]
\begin{center}
\includegraphics[width=8.0cm]{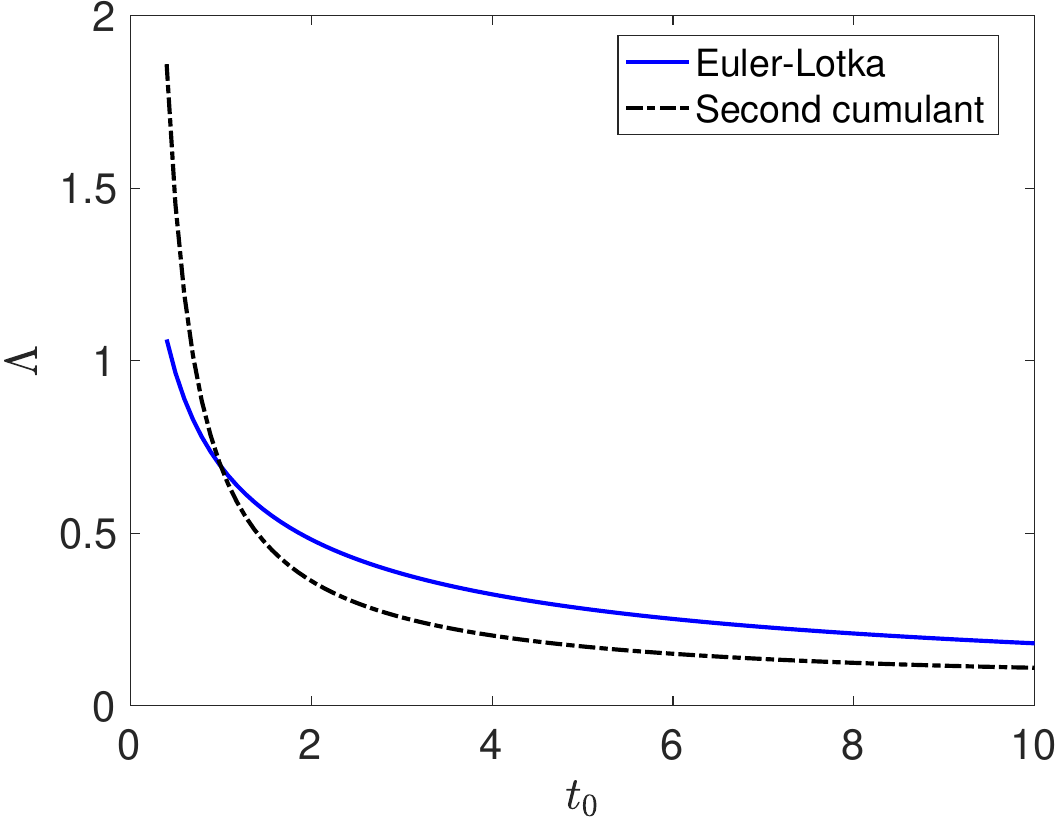}
\caption{A counter example where the SCE fails. The (exact) result of the Euler-Lotka equation is given by the solid blue line (Eq.~\eqref{recursive}), and is compared with the SCE method (black, dashed line), which relies only on the first and second cumulants of the distribution (Eq.~\eqref{SCEbad}).}
\label{Fi:Fig7}
\end{center} 
\end{figure}

The predictions of Eqs.~\eqref{recursive} and \eqref{SCEbad} are compared in Fig.~\ref{Fi:Fig7}, showing that the second cumulant method is inaccurate. Indeed, even for the case of independent generation times, the (exact) Euler-Lotka equation tells us that knowledge of the entire generation time distribution is needed; hence, the mean and variance are insufficient - therefore, the SCE cannot be guaranteed to yield accurate results for non-Gaussian generation time distributions.

\section{Connection to other work}\label{sec:otherwork}

\subsection{Jarzynski's Equality estimators}\label{sec:jarzynski}
Here, we discuss in greater detail the connection to Jarzynski's Equality estimators of free energy differences, with particular emphasis on the results presented in \cite{suarez2012phase,palassini2011improving}. These studies examine a physical system parameterized by the system size $n$, which is driven out of equilibrium between two configurations, $\Gamma_1$ and $\Gamma_2$. The well-known Jarzynski's Equality relates the work performed during this process, $W$, to the free energy difference between the two configurations, $\Delta F$, through the expression
\begin{equation}\label{eq:jarzynski}
\Delta F = - \frac{1}{\beta} \ln \mathbb{E}[e^{-\beta W}].
\end{equation}

Using Equation \ref{eq:jarzynski}, the free energy difference can be estimated by substituting the expectation with the empirical average:
\begin{equation}
\widehat{\Delta F} = -\frac{1}{\beta} \ln \left\langle e^{-\beta W^{(i)}}\right\rangle.
\end{equation}
The connection to growth rate estimation becomes evident in this context.

In \cite{suarez2012phase}, the authors explain that for small $M$, $\widehat{\Delta F}$ is a biased estimator of $\Delta F$ and analyze this bias by relating the quenched average $\mathbb{E}[\widehat{\Delta F}]$ to the free energy of the REM, obtaining results very similar to ours for the bias. 

Interestingly, they also derive finite size corrections, which would correspond to finite time corrections to the finite lineage bias in our case. These come from earlier work on finite size correction to the REM. It appears that these corrections are less relevant in our application, since our data is in the high temperature regime of the REM.

\subsection{ATM networks}

Estimating large deviation rate functions also plays a significant role in admission control for Asynchronous Transfer Mode (ATM) networks. For a detailed discussion of this application, we refer to \cite{lewis1998practical}. The basic setup is as follows: The goal is to process an input stream (e.g., service requests) where arrivals occur at stochastic times and are serviced at a constant rate. These requests are queued in a buffer of length $b$, and once the buffer is full, additional calls are discarded. The objective is to optimize the service rate to minimize or prevent call loss.

In \cite{lewis1998practical}, it is shown that this problem can be reduced to estimating the large deviation rate function $I$ of the process $A_t$, which counts the arrivals of requests in the queue. Specifically, the tail probability of the queue length $Q$ is given by
\begin{equation}
P(Q > q) \sim e^{-q \delta},
\end{equation}
where
\begin{equation}
\delta = \min\left\{\frac{I(a)}{a} : a \geq 0\right\}.
\end{equation}
Thus, $\delta$ can be derived from an estimate of the Scaled Cumulant Generating Function (SCGF) of $A_t$. This connection has motivated rigorous analysis of the convergence properties of SCGF estimators.

To our knowledge, the interplay between the fixed time and fixed count ensembles has not been explored in this context.

\section{derivation of finite-time growth rate for FDE}
\label{ap1}
We can compute the finite-time population growth rate for FDE by making the approximation that the lineage distribution is multivariate Normal, that is, 
\begin{equation}
p_{\rm lin}(\btau) \approx \frac{e^{-(\btau - \bar{\btau})^T K_{n}^{-1}(\btau - \bar{\btau})}}{(2\pi)^{n/2}{\rm det}K_{n}^{1/2}}.
\end{equation}
Here, $K_{n}$ is the matrix with entries $K_{n,i,j} = {\rm cov}(\tau_i,\tau_j)$ and $\bar{\btau} = {\mathbb E}[ \btau] =(\bar{\tau},\cdots, \bar{\tau})^T$. This formula will be exact for any Gaussian process, e.g., any autoregressive process. In particular, for an ${\rm AR}(1)$ process,
\begin{equation}
\tau_{n+1} = \bar{\tau}(1-c) + c \tau_n + \eta_n
\end{equation}
 we would have  $K_{n,i,j} = \sigma_{\eta}^2/(1-c^2)c^{|i-j|}$.
 
 Using the well-known formula for the moment generating function of a multivariate Gaussian (see e.g. \cite{rasmussen2004}), 
\begin{equation}\label{mgf}
{\mathbb E}\left[ e^{-\Lambda {\bf 1}^T\btau }\right] = e^{-\Lambda {\bf 1}^T\left(\bar{\btau} - \frac{\Lambda}{2}K_{n}{\bf 1}\right)} 
=  e^{-\Lambda n \bar{\tau} + \frac{\Lambda^2}{2}{\bf 1}^TK_{n}{\bf 1}}
\end{equation}

 In the special case of an ${\rm AR}(1)$ process with $\sigma_{\xi}^2 = (1-c^2)\sigma_{\tau}^2$, we get
\begin{equation}
\label{var}
\frac{\mathbf{1}^{T}\mathbf{K_{n}}\mathbf{1}}{n}=-\frac{\left(n  \left(c ^2-1\right)-2 c  \left(c ^{n }-1\right)\right) \sigma _{\tau }^2}{n (c
   -1)^2}
\end{equation}
where the approximation neglects terms exponentially small in $c$. Using \eqref{var}, for the variance of the FDE estimator, we now have to solve the following equation to solve the FDE equation:
\begin{equation}
 \frac{\Lambda^2}{2n}\mathbf{1}^{T}\mathbf{K_{n}}\mathbf{1}-\Lambda\tau_0+\ln(2)=0~.
 \end{equation}

Solving for $\Lambda$ and multiplying the top and bottom by 
\begin{equation}
\tau_{0}+\sqrt{\tau_{0}^2+\frac{2(1+c)\sigma_{\tau}^2\ln(2)}{1-c}}~
\end{equation}
gives the $n$-dependent growth rate to be

\begin{equation}
\label{final}
\Lambda_\text{n}=\frac{2\ln(2)/\tau_{0}}{1+\sqrt{1-\frac{\sigma_{\tau}^2\left((1-c^2)2\ln(2)n+4c(c^{n}-1)\ln(2)\right)}{(1-c)^2\tau_{0}^2n}}}~.
\end{equation}
We can write Eq.~\ref{final} in terms of the asymptotic solution as $n$ goes to infinity,
\begin{equation}
\label{ftconvergence}
    \Lambda_{\text{n}}=\Lambda+\frac{B}{n}~,
\end{equation}
where $\Lambda$ is shown in Eq.~\eqref{eq:FDE_c}, and

\begin{equation}
\label{exactcoeff}
\begin{split}
B = & \frac{4 c \ln(2)^2\sigma_\tau^2\sqrt{1-2\ln(2)\frac{\sigma^2_\tau}{\tau_{0}^2}\frac{1+c}{1-c}}}{\left(c-1\right)\tau_{0}\left(\left(c-1\right)\tau_{0}^2 + 2\sigma^2_\tau\ln(2)\left(1+c\right)\right)} \\
& \times \frac{1}{\left(1+\sqrt{1-2\ln(2)\frac{\sigma^2_\tau}{\tau_{0}^2}\frac{1+c}{1-c}}\right)}.
\end{split}
\end{equation}

\section{The mapping of FTE to REM}
\label{REM_A}
In the FTE, we can similarly make the connection to the REM by viewing $\ln(2) N_t^{(i)}$ as the scaled energy levels. Since $N_t$ is a counting variable, Eq.~\eqref{eq:fed_limit} is no longer valid. The REM with discrete energy levels has been studied in Refs.~\cite{moukarzel1992rem,ogure2009analyticity,derrida2015finite}, where both binomial and Poisson energy distributions have been studied. We found that a Gaussian approximation nevertheless seems to capture the convergence very well in the regimes we are interested in.

In our earlier work \cite{levien2020large}, we derived the rate function for the autoregressive generation time model:
\begin{equation}
I(y) = \frac{(1-c)^2}{2 \sigma_{\tau}^2}(\tau_0 - 1/y)^2.
\end{equation}
A Taylor expansion around $y = 1/\tau_0$ yields a Gaussian approximation:
\begin{equation}
\ln(2)N_t \approx t\ln(2)/\tau_0 + \ln(2)\sqrt{t}\sigma_y X_i.
\end{equation}
This motivates the following definitions:
\begin{equation}
\alpha_{\rm FTE} = \sqrt{\frac{2t}{N}},
\end{equation}
and
\begin{equation}
\beta_{\rm FTE} = \ln(2)\sigma_y\alpha_{\rm FTE}.
\end{equation}
where $\sigma_y =  \sigma_{\tau}^2/(1-c)^2$. 
Similar to the approach taken for the FDE, we can rewrite Eq.~\eqref{eq:fte_scgf} in terms of \(f_N(\beta)\):
\begin{equation}
\tilde{\Lambda}_{\rm REM}(\alpha) = \frac{\ln(2)}{\tau_0} + \frac{1}{\alpha^2}\left(-\beta f_N(\beta) - \ln(2)\right).
\end{equation}
Once we replace $f_N(\beta)$ with $\bar{f}_N(\beta)$ we obtain 
\begin{equation}\label{eq:Lhat_rem2}
\tilde{\Lambda}_{\rm REM}(\alpha)=  \left\{ \begin{array}{lr}
 \tilde{\Lambda} & \alpha < \alpha_c\\
\frac{\ln(2)}{\tau_0} + \frac{1}{\alpha^2}(\sqrt{\ln(2)}-\ln(2)) &\alpha \ge \alpha_c
 \end{array}\right.
\end{equation}
where $\tilde{\Lambda} = \ln(2)/\tau_0 + \ln(2)^2\sigma_y^2/4$, which is the Taylor expansion of $\Lambda$ given by Eq.~\ref{eq:FDE_c}
in $\sigma_y^2$. Note that the transition occurs at the same critical $\alpha$ for both estimators but the decay to the naive solution of $\ln(2)/\tau_0$ is different.

In Fig. 2 and in Ref.~\cite{levien2020large}, we used $\sigma=0.2$. For this value of the noise, $\alpha_c\approx 12$, and $n< 300$ is needed to avoid the linearization effect. This analysis qualitatively agrees with a more heuristic approach in Ref.~\cite{levien2020large} where we obtained the criteria $n<\frac{\ln(M)}{\sigma^2\ln(2)}\approx100$ divisions.

\section{Fits of the experimental data}
\label{modelfit}
Simulated data was generated by fitting autoregressive models to the experimental \emph{E. coli} data. First, we fit the autoregressive model given by Eq. \eqref{modeleq1}. This was achieved by performing a simple linear regression. Next, in order to account for the fact that cell-size is regulated, we fit a multivariate autoregressive model, where log cell length was included as a predictor. This takes the form 
\begin{align}
\label{model1}
\bx_{i+1} = A\bx_i + {\bf b} + {\boldsymbol{\xi}_i}
\end{align}
Here, $A\in \reals^{2 \times 2}$ and ${\bf b}\in \reals^2$ are coefficients to be fitted and ${\boldsymbol{\xi}_i} \in \reals^2$ is a noise vector. The regression variable is
\begin{equation}
\label{model2}
\bx_i = \left[\begin{array}{c} 
\ln s_i\\
\tau_i
\end{array}\right]
\end{equation}
and $s_i$ is the size of the $i$th cell at birth. 

We used standard least squares for regression with multiple response variables \cite{demidenko2019advanced} to fit the coefficients $A$ and ${\bf b}$, as well as the noise magnitudes.

Note that by simply including the additional variable of log cell-size, cell-size is automatically regulated. We could have alternatively included growth rates, instead of generation time and/or log fold change in size as predictors. This was the approach taken in \cite{kohram2021bacterial}. However, we found that for the purpose of predicting growth rate and the convergence pattern of the FTE and FDE, simply adding the additional predictor of size was sufficient.

\begin{figure}[h!]
\begin{center}
\includegraphics[width=8.8cm]{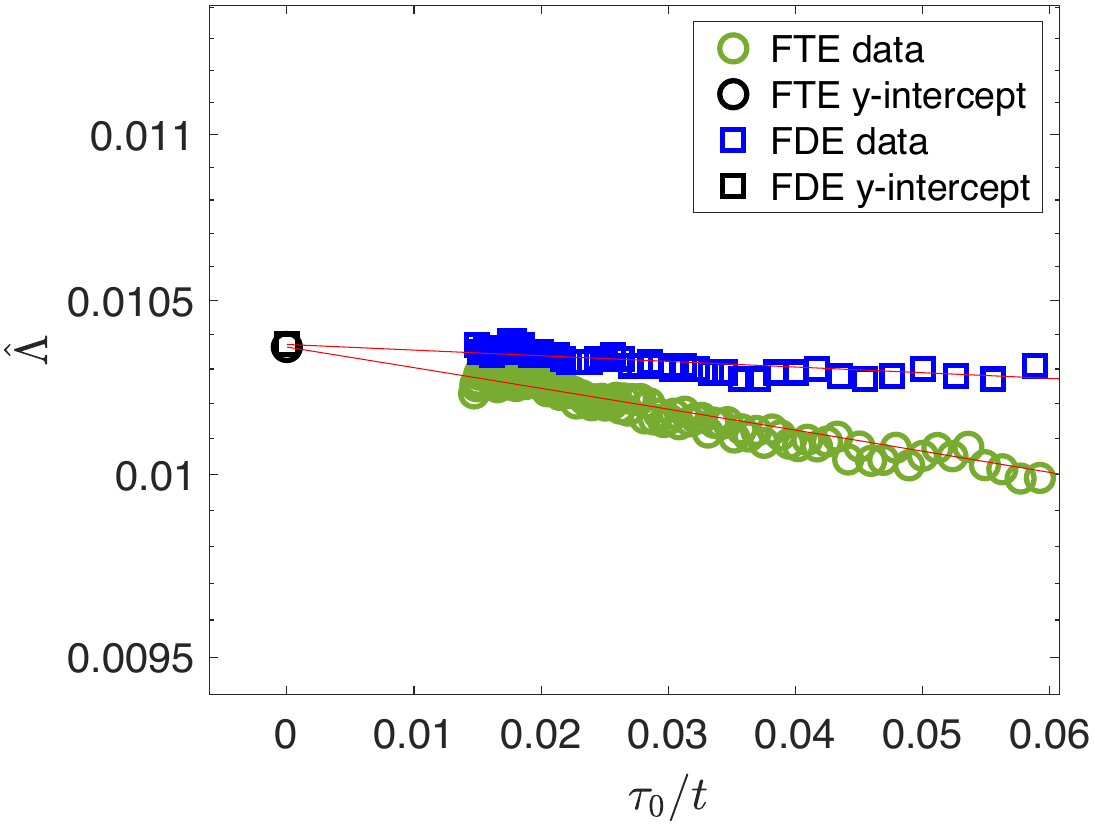}
\caption{Estimates of $\Lambda$ using the FDE and FTE with finite size corrections. By plotting as a function of $\tau_0/t$, we can fit the convergence to a line. The symbols represent the FDE (blue squares) and FTE (green circles) applied with no corrections. The solid red solid lines represent the fit and the black symbols are y-intercepts which give the long-time population growth rate.}
\label{Fi:SI1}
\end{center} 
\end{figure}

\end{document}